\documentclass[12pt]{article}

\usepackage{amsmath}
\usepackage{amssymb}
\usepackage{amsfonts}
\usepackage{amscd}
\usepackage{amsbsy}
\usepackage{amsthm}
\usepackage{latexsym}
\usepackage{graphicx}
\usepackage{slashed}
\usepackage{color}

\DeclareMathOperator{\cx}{\square}

\def\beq{\begin{eqnarray}}
\def\eeq{\end{eqnarray}}
\newcommand{\nn}{\nonumber}

\def\al{\alpha}
\def\be{\beta}

\def\ga{\gamma}
\def\de{\delta}
\def\vp{\varepsilon}
\def\ep{\epsilon}
\def\ze{\zeta}

\def\ka{\kappa}
\def\la{\lambda}
\def\na{\nabla}
\def\pa{\partial}

\def\si{\sigma}
\def\om{\omega}

\def\Ga{\Gamma}

\def\La{\Lambda}
\def\Si{\Sigma}

\usepackage[symbol]{footmisc} 
\usepackage{float} 
\usepackage{cite}  
\usepackage{titlesec}

\titleformat*{\section}{\large\bfseries}
\titleformat*{\subsection}{\normalsize\bfseries}

\begin{document}

\begin{center}

{\large\bf
Decoupling theorem and effective quantum gravity
}
\vskip 5mm

\textbf{Ilya L. Shapiro}
\footnote{E-mail address: \ ilyashapiro2003@ufjf.br}
\footnote{On leave from Tomsk State Pedagogical University.}
\footnote{Invited contribution to the volume in memoria of Vladislav
Gavrilovich Bagrov.}

\vskip 5mm

{\sl Departamento de F\'{\i}sica, ICE,
Universidade Federal de Juiz de Fora
\\
Campus Universit\'{a}rio - Juiz de Fora, 36036-900, MG, Brazil}
\end{center}

\vskip 6mm

\centerline{\textbf{Abstract}}
\vskip 1mm

\begin{quotation}
\noindent
This is a contribution to the memorial edition devoted to Professor
Vladislav Gavrilovich Bagrov, who was my official adviser from the
beginning of undergraduate period to the end of Ph.D. The text
includes a mentioning of two my publications in Izvestia VUZov
Fisica (Russian Physics Journal), where Vladislav Gavrilovich
served as an Editor.

The rest of this paper is based on the recent lectures about
decoupling in quantum gravity, given in ICTP-SAIFR in S\~ao
Paulo and at the school ``Estate Quantistica'' in Scalea, Italy.
After a brief and mainly qualitative review of the decoupling theorem
in semiclassical gravity and the scalar fourth-derivative model of
Antoniadis and Mottola, we explain what is the expected result for
the physical beta functions in fourth-derivative quantum gravity and
what should remain from these beta functions in the IR.
\vskip 3mm

\vskip 2mm


\end{quotation}

\section{My interactions with V.G. Bagrov
and the Russian Physics Journal}
\label{Remi}

I went to ask a supervision of V.G. Bagrov at the beginning of
the second year in the University, after a serious inspection of
all possibilities offered by Tomsk State University at that time.
Almost fifty years after this, my feeling is that I was a lucky
man. It was always nice to communicate with Vladislav Gavrilovich,
I became a member of selected group of students, got many useful
advises from V.G. Bagrov and, 15 years after he accepted
me as a student, we even published a paper together. The journal
Izvestia VUZov Fisica (nowadays Russian Physics Journal) was
an important ingredient of our scientific life at that time, in
particular because it was necessary to publish the paper in
Russian before sending it to Physics Letters. On the other hand,
I was not only an author, but also a referee. Along with assisting
in the oral exams in Mathematical Physics, writing reports for
Izvestia VUZov was our important duty and, years after, now
I believe that it was a useful experience. Sometimes, the papers
which I had to report on where interesting and other looked very
strange, to say the least. But for a young scholar all experiences
serve for good, so I cannot complain.

My first paper published in this journal was \cite{1storder}.
At that moment we (my main adviser I.L. Buchbinder and I)
just learned about the background field method and this paper
was my very first application of this fresh knowledge to the quantum
gravity in first order formalism. For me, this work was a remarkable
point because this method became one of the standard technologies
which we used in future works. Curiously enough, the calculations
done in this our paper were redone by other authors decades later.

The second work which I would like to mention was our common
paper with Buchbinder and Bagrov \cite{BBSh}. This one was
about the possible low-energy manifestations of torsion as a
background for a Dirac field. The English version was rejected
by Phys. Rev. D who argued towards ``torsion is not Physics''.
Three decades after, it is satisfactory to observe that this great
journal changed its politics and published many works on the same
subject. So, we can see that sometimes our Izvestia VUZov was
gaining in the scientific competition.

\section{Introduction}
\label{Intro}

The simplest possible theory of four-dimensional quantum
gravity (QG) is based on GR, but it does not pass the test of
renormalizability. That means loop corrections contribute to
the qualitatively new terms in the action of gravity, making
its particle spectrum to become different with every loop. There
is no known consistent way to resolve this problem, but one of
the most natural solutions is to start with the higher derivative
theory providing renormalizability \cite{Stelle77} or even
superrenormalizability \cite{highderi}. The common point
of all such models is the presence of extra degrees of freedom
in the spectrum, along with the massless graviton of GR.

Decoupling is an important concept in QFT and many applications
ranging from particle physics to cosmology. The notion of quantum
decoupling looks especially relevant in QG, where it is supposed to
explain how the fundamental theory of quantum gravity behaves in
the IR (i.e., in the low-energy limit).  The purpose of this note is to
summarize the known features of decoupling and of perturbative
quantum gravity, and on this basis, explain what is the expected
result for the decoupling theorem in higher derivative QG.

Let us mention a few references. The pioneering work about quantum
decoupling is the paper where the Appelquist and Carazzone theorem
was formulated in QED \cite{AC}. A good ``standard'' review is
\cite{Manohar} and a detailed introduction can be found in the recent
textbook \cite{OUP}. The first discussion of decoupling in higher
derivative QG was in one of the appendices of the classical work of
Fradkin and Tseytlin \cite{frts82}. The first explicit calculation of
decoupling with the gravitational background was in \cite{apco}.

The rest of the article is organized as follows.
Sec.~\ref{sec2} includes a qualitative review of the notion of
decoupling, including the recent work about the fourth-derivative
toy model \cite{EffAM}. Sec.~\ref{sec3} explains the critical
difference between the Antoniadis and Mottola scalar model
and a real QG theory with four derivatives \cite{Stelle77,frts82}.
We summarize the existing information about loop corrections
and explain what should be expected from the complete
analysis of the IR decoupling and the physical scheme of
renormalization in this case. Finally, in Sect.~\ref{Conc},
we draw our conclusions.

\section{Decoupling theorem}
 \label{sec2}

The particle contents of a QG model are defined by the structure
of the propagator and the last depends on up to the second-order
in curvature terms,
\beq
S_{gen}
&=&
\int_x
\,\Big\{ - \frac{1}{\ka^2} \,(R+2\La)
\,+\, \frac12\,C_{\mu\nu\al\be} \,\Phi(\Box)\,C_{\mu\nu\al\be}
\nn
\\
&&
\quad
+ \,\,
\frac12\,R \, \Psi (\Box)\, R
\,+\, {\mathcal O}(R_{\dots}^3)\Big\}\,.
\nn
\eeq
where $\int_x = \int d^4 x \sqrt{-g}$. \
For the polynomial functions $\,\Phi(\Box)$ and
$\,\Psi(\Box)$ there is always a massless graviton and also massive
degrees of freedom, which can be ghosts, tachyons,
tachyonic ghosts, or normal particles.

The one-loop contributions of matter fields with mass $m$ to the
form factors $\Phi(\square)$ and $\Psi(\square)$ derived in
\cite{apco} (see \cite{OUP} for other relevant references)
has shown that these form factors are ``controlled'' by the
logarithmic divergences in the UV and demonstrate quadratic
$\mathcal{O}(\square/m^2)$ decoupling in the IR. In all these
cases, the result was coming from a loop with internal lines of
massive degrees of freedom.

To understand the decoupling in the quantum gravity model with
massive degrees of freedom, one has to work out also the mixed
loops, that is those with some of internal lines being of massless
and another of the massive degrees of freedom. What happens with
these modes in the IR? Is there a decoupling?
The main hypothesis of effective QG is that the IR limit in a QG
model is always the quantum version of GR \cite{don}. However,
this output is not granted, as there may be other possibilities
\cite{Polemic}.

There are no doubts that the calculations of the form factors in
 higher derivative QG would be a interesting work to do. However,
in this case one needs to deal with a complex theory, with scalar
and tensor modes, gauge fixing, Faddeev-Popov ghosts, the third
ghost, etc. In the recent work
\cite{EffAM}, the behavior of mixed loops was explored in a
simplified toy model of the QG theory \cite{antmot}, with
qualitatively similar structures of propagator and vertices. The
similarity of these theories led to the original proposal for this
model \cite{OdSh-91}.

The classical action of the model results from the integration of
trace anomaly,
\beq
\label{action-AM}
S_{\textrm{cf}} \, = \, \int d^4 x
\Bigl\{
\ga e^{2\si}(\pa\si)^2  - \la e^{4\si}
-\theta^2 (\square\si)^2
- \ze\left[2(\pa\si)^{2}\square\si
+ (\pa\si)^{4}\right]\Bigr\},
\eeq
where $(\pa\si)^ 2 = \eta^{\mu\nu}\pa_\mu \si \pa_\nu \si$ and
the coefficients $\theta^2$ and $\ze$ depend on the particle
contents of the original theory  of conformal matter fields,
$\ga = 3/8\pi G$ and $\la = \La/8\pi G$.
The IR sector includes Einstein-Hilbert and cosmological terms
for a conformally flat metric,
\beq
\label{action-IR}
S_{\textrm{IR}} \, = \, \int d^4 x
\Bigl\{\ga e^{2\si}(\pa\si)^{2}-\la e^{4\si} \Bigl\}.
\eeq

Using the standard algorithm \cite{frts82,bavi85} for the
fourth-order operators, we get the divergences in the ``fundamental''
four-derivative model \cite{antmot}
\beq
\label{div-AM}
&&
\bar{\Ga}_{\textrm{div}}^{(1)}
\,=\,
-\, \frac{1}{\vp}
\int d^4 x
\Big\{\frac{5\ze^{2}}{\theta^4}
\left[\square\si + (\pa\si)^{2}\right]^2
+\frac{\ga}{\theta^2}\Big(\frac{3\ze}{\theta^2}
+2\Big)(\pa\si)^{2}e^{2\si}
\\
\label{1}
&&
\qquad \qquad
- \,\, \Big(\frac{8\la}{\theta^2}
- \frac{\ga^{2}}{2\theta^4}\Big) e^{4\si}\Big\},
\quad
\mbox{where}
\quad
\vp=(4\pi)^{2}(n-4).
\eeq
This expression fits the UV limit of the form factors,
providing correspondence between Minimal Subtraction and
(physical) Momentum Subtraction renormalization schemes.

The divergences of the effective theory (\ref{action-IR}) have
the form
\beq
\label{div_IR}
\bar{\Ga}_{\textrm{div,\,IR}}^{(1)}=-\frac{1}{\vp}\int
d^{4}x\,\bigg\{\frac{1}{2}\big[\square\si
+(\pa\si)^{2}\big]^{2}
-\frac83\La\,e^{2\si}(\pa\si)^{2}
+\frac{32}{9}\La^{2}e^{4\si}\bigg\}.
\eeq
In agreement with the power counting, the fourth-derivative
counterterms emerge because the theory is non-renormalizable.

Let us briefly summarize the results in the Momentum Subtraction
scheme \cite{EffAM}. Similar to the four-derivative QG, the
Antoniadis and Mottola model has a light mode and heavy mode
propagating. The masses of these modes are
\beq
m^2 = \frac{8\La}{3},
\qquad
M^2=\frac{\ga}{\theta^2},
\qquad
m \ll M\,.
\label{masses}
\eeq
Without the cosmological constant, the light mode is massless.
The basic elements of the relevant Feynman technique
include the propagator of the light mode (simple line),
 of the large-mass mode (double line), and all types of vertices.

At the tree level, the heavy mode collapses in the IR, such
that there is only the light field propagating. The question is
what happens in the IR with the loop corrections?
In the self-energy part, there are three types of the diagrams,
but only the bubble-type ones, shown in Fig. 1, contribute to
the nonlocal form factors.
\begin{figure}[!ht]
\centering
\includegraphics[scale=0.80]{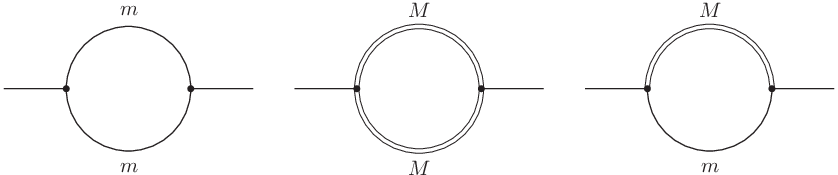}
\begin{quotation}
\textit{Fig. 1. \ Three types of diagrams of the bubble type.}
\end{quotation}
\end{figure}

The first type of loop produces the form factors which are pure
logarithms. That is also true in QG, for both quantum metric and
the Faddeev-Popov ghost sectors.
The diagrams of the second type produce a more complicated form
factors  \cite{apco,OUP}, demonstrating usual quadratic decoupling,
according to the Appelquist and Carazzone theorem.
The remaining question is what happens in the IR with the
contributions of the third type of diagrams. Other sorts of diagrams
are shown in Fig. 2. These diagrams are relevant in the sense they
produce logarithmic divergences, the same as bubble diagrams.
However, they do not contribute to the nonlocal form factors.
\\
\begin{figure}[!ht]
\centering
\includegraphics[scale=0.78]{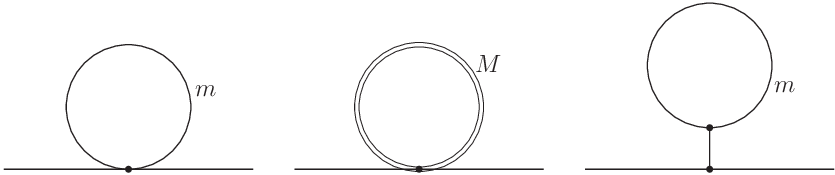}
\begin{quotation}
\noindent
\textit{Fig. 2. \ Examples of other types of diagrams:
snail and tadpole.}
\end{quotation}
\end{figure}

The main result for the nonlocal form factors can be expressed
via the self-energy functions $\Si$. The UV limits of these
expressions faithfully repeat the structure of the logarithmic
divergences. The IR limit means assuming $m^2 = 0$
(this is just for simplification)
and $M^2\gg p^2$. Even in this case, the results of the calculations
of the form factors are very cumbersome \cite{EffAM} and
we skip all the details.
Qualitatively, the result can be summarized in Fig. 3.
\\
\begin{figure}[!ht]
\centering
\includegraphics[width=9.2cm,angle=0]{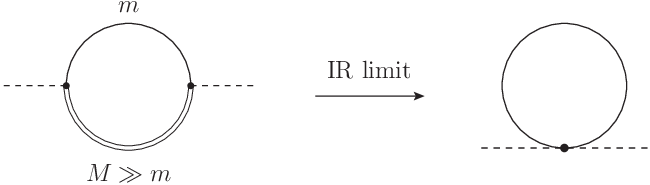}
\begin{quotation}
\noindent
\textit{Fig. 3. \ The mixed bubble diagram collapses into a snail
diagram in the IR limit. In the IR, this diagram
does not contribute to the nonlocal form factor.}
\end{quotation}
\end{figure}

Only the contributions of the massless fields remain present in
the IR limit. One can verify \cite{EffAM} that there is a perfect
correspondence between the IR contributions of the fundamental
theory (\ref{action-AM}) and the UV contributions, or simply the
logarithmic divergence (\ref{div_IR})
in the effective theory (\ref{action-IR}).
Since the features of this toy model are qualitatively the same as
the ones of the fourth derivative QG, we can trust that the
decoupling of massive and mixed loops in QG occurs in the
same way. Let us see what we can expect, under this assumption,
as the IR limit of the fourth derivative QG.

\section{What should we get in quantum gravity}
\label{sec3}

To understand the main difference between the two models, let us
remember the well-known facts about the gauge and parametrization
dependencies of loop corrections in QG.

The UV ``running'' depends on divergences and the last may be not
universal. Indeed, there may be dependence on gauge-fixing and
parametrization. Typically,
\beq
S_t(\al_i)\,=\,S_{QG}+ S_{gf}+S_{ghost}\,,
\label{FP}
\eeq
where
$\al_i = (\be_k,\,\,\ga_j)$ and parameters used in the gauge fixing
and parametrization of quantum fields. For instance, the gauge fixing
action is
\beq
&&
S_{gf} = 
\int_x
 \, \, \chi_\mu \,Y^{\mu\nu}(\be_2,\,\be_3,\,...)\,\chi_\nu
,\qquad
\chi_\mu = \na_\lambda\,\phi^\lambda_\mu
- \be_1 \,\na_\mu \phi^{\lambda}_{\lambda}\,.
\label{gf}
\eeq
The general one-loop parametrization of quantum metric
$\phi_{\mu\nu}$ is \cite{JDG-QG}
\beq
\nn
g_{\mu\nu}
&
\,\,\longrightarrow \,\, & g'_{\mu\nu}
\,=\,
g_{\mu\nu}
+ \ka \big(\ga_1\, \phi_{\mu\nu}
+ \ga_2\, \phi\, g_{\mu\nu} \big)
\nonumber
\\
&&
\,+\,\ka^2 \big(\ga_3\, \phi_{\mu\rho}\phi^\rho_\nu
+ \ga_4\, \phi_{\rho\om} \phi^{\rho\om} \,g_{\mu\nu}
+ \ga_5\,\phi \, \phi_{\mu\nu}
+ \ga_6\, \phi^2 \,g_{\mu\nu}
\big),
\label{param}
\eeq
where
$g_{\mu\nu}$
is the background metric.

We can use the general theorems about gauge-fixing and
parametrization independence of the effective action on-shell
\cite{aref,VLT82} (see more references in \cite{OUP}).
The difference between the divergences 
of the one-loop effective action, evaluated using different gauge
and parametrization parameters $\alpha_i$ and $\alpha_0$
is proportional to the classical equations of motion
\beq
&&
\de \bar{\Ga}_{div}^{(1)} \,=\,
\bar{\Ga}_{div}^{(1)} (\alpha_i)
- \bar{\Ga}_{div}^{(1)} (\alpha_0)
\,=\, \frac{1}{\ep}
\int_x \,  f_{\mu\nu}\, \vp^{\mu\nu},
\qquad
\label{amb-GR}
\\
&&
\mbox{where, in pure GR,}
\quad
\vp^{\mu\nu}
= R^{\mu\nu} - \frac12 \,g^{\mu\nu} (R + 2 \La).
\qquad\qquad\qquad
\nn
\eeq
In QG based on GR one meets the divergences
\beq
\Ga^{(1)}_{div} = \frac{1}{\ep}\,
\int_x 
\Big\{
c_1\,R_{\mu\nu\al\be}^2 + c_2R_{\al\be}^2 + c_3R^2
+ c_4  {\Box}R + c_5R + c_6 \Big \}.
\label{divsGR}
\\
\mbox{and}
\qquad
\vp^{\mu\nu}\,=\,\frac{1}{\sqrt{-g}}\,\frac{\de S}{\de g_{\mu\nu}}
\,=\,R^{\mu\nu}-\frac12\,\big(R+2\La\big)g^{\mu\nu}\,.
\label{eqmo}
\eeq
Therefore, in this case
\beq
f_{\mu\nu}\,=\,
b_1 R_{\mu\nu} + b_2 Rg_{\mu\nu} + b_3  \La g_{\mu\nu}
+ b_4 g_{\mu\nu}\Box + b_5 \na_\mu \na_\nu
\label{fmn}
\eeq
and we arrive at
\beq
&&
\de \Ga^{(1)}_{div}
\,=\,
\Ga^{(1)}_{div}(\al_i) \,-\, \Ga^{(1)}_{div}(\al^{0}_i)
\nn
\\
&&
\,=\, \frac{1}{\ep}\,
\int_x 
\big(
b_1 R_{\mu\nu} + b_2 Rg_{\mu\nu} + b_3  \La g_{\mu\nu}
+ b_4 g_{\mu\nu}\Box + b_5 \na_\mu \na_\nu\big) \,\vp^{\mu\nu},
\label{b12345}
\eeq
where the parameters
 \ $b_{1,2,..,5}$  \
depend on the full set of 
\ {$\al_i$}. Using this formula, one can easily establish the
ambiguities in the coefficients in (\ref{divsGR}).
The universal (invariant) beta functions include the one for the
coefficient of the Gauss-Bonnet term and the unique combination
of other coefficients \cite{frts82,JDG-QG}
\beq
c_1
\qquad
\mbox{and}
\qquad
c_{inv}\,=\,c_6 - 4\La c_5 + 4\La^2 c_2 + 16\La^2 c_3\,.
\label{invars}
\eeq
Thus, in quantum GR, there are
no real beta functions for the $C^2$ and $R^2$ terms.

This conclusion does not apply to other models of QG.
In the fourth-derivative QG, local divergences have the form
\beq
\Ga^{(1)}_{div} = \frac{1}{\ep}\,
\int_x 
\big\{
c_1\,R_{\mu\nu\al\be}^2 + c_2R_{\al\be}^2 + c_3R^2
+ c_4  {\Box}R + c_5R + c_6 \big \}.
\label{4derdivs}
\eeq
In this case, the classical equations of motion are also four-derivative.
Therefore,
\beq
&&
\de \bar{\Ga}_{div}^{(1)} \,=\,
\bar{\Ga}_{div}^{(1)} (\alpha_i) - \bar{\Ga}_{div}^{(1)} (\alpha_0)
\,=\, - \, \frac{1}{\ep}
\int_x \, \vp^{\mu\nu} f_{\mu\nu},
\label{ambi-HDQG}
\\
&&
\mbox{with}
\qquad
f_{\mu\nu} ({\alpha}_{i})
\,=\, g_{\mu\nu} \,f(\alpha_i).
\label{fmn-HDQG}
\eeq
Here
$f({\alpha}_{i})$ is an arbitrary dimensionless function of
the parameters $\alpha_i$.
The gauge dependence of the divergent part
is controlled by the ``conformal shift'' of the classical action
\beq
\Ga^{(1)}_{div}({\alpha}_{i})
\,-\,
\Ga^{(1)}_{div}(\al^{0}_i)
\,\,=\,\,
f({\alpha}_{i}) \int d^4 x \, \,
g_{\mu\nu}\,\frac{\delta S}{\delta g_{\mu\nu}}.
\label{confshift}
\eeq
In the conformal model of QG, the {\it r.h.s.} vanishes owing to
the Noether identity for this symmetry. In the general theory, this
isn't the case. In this way, we arrive at
\beq
\Ga^{(1)}_{div}({\alpha}_{i})
\,-\,
\Ga^{(1)}_{div}(\al^{0}_i)
\,\,=\,\,
f(\alpha_{i}) \int_x 
\,\Big\{ \frac{2\omega}{\lambda}\cx R
\,-\,\frac{1}{\ka^2}(R+4\Lambda)\Big\}.
\mbox{\qquad}
\label{ambi-f-HDQG}
\eeq
The three divergent coefficients depend on the gauge fixing and
parametrization. However, there is an invariant combination of the
coefficients of Einstein-Hilbert and cosmological terms. In the
fourth-derivative QG, there is a well-defined UV running of the
coefficients of the Gauss-Bonnet, $C^2$ and $R^2$ terms and an
invariant combination of the beta functions for the Newton constant
and cosmological constant.

Let us compare  UV running in quantum GR and fourth-derivative
QG. In quantum GR, there is only one well-defined renormalization
group equation on shell \cite{frts82}.
\beq
\label{onshellRG}
\mu\,\frac{d\ga}{d\mu}
\,=\, 
\,-\,\frac{29\,\,}{5(4\pi)^2} \,\ga^2,
\qquad
\ga = 16\pi G \La.
\eeq
However, this cannot be seen as an indication of asymptotic freedom,
as it is on shell.

In the fourth derivative QG
\beq
S\, = \,-\,\int_x \,
\Big\{\frac{1}{2\la}\,C^2 \,-\,\frac{1}{\rho}\,E_4
\,+\,\frac{1}{\xi}\, R^2 \, - \,
\frac{1}{\kappa^2}\, (R - 2 \La )\,\Big\}
\label{action-HDQG}
\eeq
 there are well-defined renormalization group equations,
\beq
(4\pi)^2\,\frac{d\rho}{dt} &=&
-\,\frac{196}{45}\,\rho^2\,,\quad (4\pi)^2\,\frac{d\la}{dt}
= -\,\frac{133}{10}\,\la^2\,, \nonumber
\\
\nonumber
\\
(4\pi)^2\,\frac{d\xi}{dt}
&=& -\,10\,\la^2\,\xi^2 + 5\,\la\,\xi - \frac{5}{36}\,.
\label{RG-HDQG}
\eeq
In this case, we certainly have asymptotic freedom for the
effective charge $\la$ and the UV stable fixed point for the
ratio $\xi/\la$ \cite{frts82,avbar86}
(with several subsequent verifications).

Thus, the ambiguities in the fourth-derivative QG model and in
quantum GR  are qualitatively different. The question is what we can
expect from the decoupling in this situation? The final answer can be
provided only by explicit calculations, including at least some of the
aforementioned ambiguities. However, what can we say before doing
these difficult explicit calculations?

The main working assumptions are as follows.
\textit{i)}
The GR is a universal limit of QG in the IR. For a while, this was
checked only for a scalar Antoniadis and Mottola model and found
correct.  \textit{ii)} The decoupling of the mixed diagrams
follows the same pattern in QG and the scalar model.
If both things are true, in the UV, the masses of the extra degrees of
freedom are irrelevant and we expect a perfect fit between Minimal
Subtraction and Momentum Subtraction renormalization schemes. But
in the IR, there will be a ``mess'', in the sense that the ``beta
functions''  are ambiguous and have no physical meaning.

The unique well-defined beta function in the IR corresponds to the
running of the dimensionless ratio of the Newton constant
$G$  and the cosmological constant $\La$. Unfortunately, this is
exactly the part that cannot be explored using Feynman diagrams.
Those are certainly ``bad news'', but the positive aspect is
that we can see what we have to calculate and how we
can check the general statements in the case of effective QG.

\section{Conclusions} 
\label{Conc}

One of the most interesting aspects of the effective approach to
QG is related to the universality of the quantum GR in the IR.
In the recent paper \cite{EffAM}, we have explored the decoupling
in the fourth-derivative scalar model, which has some of the most
relevant features as the corresponding QG, e.g., massless (or light)
and massive modes, and the non-polynomial self-interactions.
This output creates positive expectations concerning the universality
of the quantum GR in the IR limit.

Assuming that the same pattern of decoupling holds in a real QG,
and using the existing knowledge of ambiguities related to gauge-
and parametrization dependence in QG, we describe what has to be
the result of an explicit calculation. Moreover, it is possible to see
which kind of calculations have to be performed to check the
expectations.


\section*{Acknowledgements}

Author is grateful to Conselho Nacional de Desenvolvimento
Cient\'{i}fico e Tecnol\'{o}gico - CNPq for the partial support
under the grant 305122/2023-1.


\end{document}